\documentclass[a4paper,11pt]{article}
\usepackage{jheppub} 
\usepackage[T1]{fontenc} 
\usepackage{hyperref}
\newcommand{\be}{\begin{equation}}
\newcommand{\ee}{\end{equation}}
\newcommand{\ba}{\begin{eqnarray}}
\newcommand{\ea}{\end{eqnarray}}
\newcommand{\ar}{\arrowvert} 
\newcommand{\ra}{\rangle} 
\newcommand{\la}{\langle} 
\newcommand{\da}{\dagger} 
 
\newcommand{\cd}{\! \cdot \!} 
\newcommand{\Dirac}{\not \!}

\title{\boldmath 
Universality of Planck's constant\ \  and a\\
constraint from the absence of $\hbar$-induced $\nu$ mixing
}

\author{Felipe J. Llanes-Estrada}
\affiliation{Departamento de F\'{\i}sica Te\'orica I, Universidad
Complutense de Madrid, 28040 Madrid, Spain}

\abstract{
You have probably often set $\hbar=1$; but for what particle?
I revisit here the possibility of a non-universal Planck-constant.
Anomaly cancellation suggests that all particles in the same family perceive the same $\hbar$ at fixed charges $e$, $g_w$, $g_s$; the difference between the muon's and the electron's (and thus the first and second families) can be tightly constrained by the muon's anomalous magnetic moment, but constraints are weaker for the third family.\\
Neutrino mixing could have proceed a priori not only by the Lagrangian neutrino mass-term, but also by the kinetic term if Planck's constant was not equal for all three species. An experimental constraint follows as such
contributions, characterized by oscillations proportional to the energy, as opposed to the inverse energy, 
have been generically analyzed in the past. This provides at the same time support for gauge invariance. \\
On the other hand if $\hbar$ differs among particles while fixing the fine structure constants $\alpha_{em}$, $\alpha_s$, etc. instead of the charges, it affects the muonic atom puzzle without much constrain from g-2.
}

\begin{document} 
\maketitle
\flushbottom
\section{Introduction}

Planck~\cite{Planck:1901} invented the constant $h$ to quantize light in discrete photon ``packages'' of energy proportional to the radiation frequency,
\be \label{photon}
E_\gamma = h \nu_\gamma \ .
\ee 
This $h$ has been measured with increasing precision in the  metric system of units ever since,
as I will briefly overview in section~\ref{sec:measureh}. 
It is customary since Bohr's time to employ angular frequency $\omega$ instead of $\nu$ in Fourier transforms and thus divide $h$ by the ubiquitous factor of $2\pi$ into Planck's reduced constant $\hbar := \frac{h}{2\pi}$,
to which I will be referring to~\footnote{The symbol seems to have been used first in this context by Dirac around 1925-1928, but it was long before employed by alchemists to denote ``Lead, dominated by Saturn'' with $\hbar$ being the divinity's scythe.}. One of its most prominent appearances is in Heisenberg's uncertainty principle for a quantum particle 
\be \label{uncertainty}
\Delta x \Delta p \ge \frac{\hbar}{2}\ .
\ee

The basic role of $\hbar$ for a material particle such as the electron is summarized by the proportionality between a  particle's momentum and the derivative of its De Broglie's matter wave,
\be \label{deBroglie}
{\bf p}_e = \hbar\ (-i \nabla \psi_e )  \ . 
\ee
This proportionality might in principle have depended on the particle species, an aspect that has not been really examined in recent literature:
Bohr's atom worked all too successfully with $\hbar$ taken from radiation work. There are no modern explicit studies similar to the empirical basis of the well tested equivalence principle for Newton-Cavendish's gravitational constant $G_N$, that seems to be the same for all material bodies and energy forms.

Customarily in atomic and subatomic physics, including particle physics, Planck's constant is absorbed into the definition of the system of units, or ``set to 1'', as is the speed of light, Boltzmann's constant and other quantities. 
Most formulae are written without the factors of $\hbar$ needed for an arbitrary system of units.
This practice is valid as long as Planck's constant is a universal constant, in the same sense of  $G_N$.
But excepting the photon and electron, for which Planck's constant has been  directly measured, the constant has been assumed to be the same for other particles without much of a direct test. The reason is probably that the foundations of quantum mechanics were established well before the particle explosion after 1945.

Planck's  constant keeps nevertheless being discussed in the theory literature. Brodsky and Hoyer~\cite{Brodsky:2010zk}, for example, have recently reformulated perturbation theory as an expansion in powers of $\hbar$ (thus needing to introduce factors of $\hbar$ back into the quantum field formalism) but have assumed that it indeed is a universal constant. 
Another aspect that has been indirectly assessed is the time variation through that of $\alpha=e^2/(4\pi\hbar)$, with negative results to date (see for example~\cite{Dzuba:1999zz}).

In this article I adopt the empirist's point of view. We do not really know what equations~\ref{photon} or~\ref{uncertainty} are about, whether a property of matter or energy, spacetime, or information. It might as well happen that such relations are slightly different for each of the elementary components of the Standard Model that we call particles (and their associated fields). I will try to constrain this possibility here. The most powerful indirect constraint follows, quite unexpectedly, for the Planck constant associated to the muon and tau neutrinos, 
$\hbar_{\nu_\mu} \simeq \hbar_{\nu_\tau}$, due to a non-standard form of neutrino mixing, as I will expose in section~\ref{sec:mixing}.

Because $\hbar$ can be shuffled between different terms of the Lagrangian by field redefinitions, as recapitulated in section~\ref{shufflehbar},  
there are at least two cases to consider in studying the variations of $\hbar$. 

One case consists of thinking that different particles perceive different Planck constants while the charges $e_i\equiv e$ are universal. This setup can be very constrained because a non-universal $\hbar$ then forces a non-universal coupling constant in some of the SM gauge interactions, with different particles having slightly different coupling constants $\alpha_{em}:=e^2/(4\pi\hbar c)$, $\alpha_s$, etc.

The mathematical consistency of the Standard Model (its ability to absorb all high-momentum quantum fluctuations in its parameters during the renormalization process) is achieved thanks to certain relations between the charges of fermions within each of the three particle families. These anomaly cancellation conditions are postponed to subsection~\ref{sec:anomalies} since they are well-known. Suffice it to say here that they are suggestive of an $\hbar$ that is the same for all particles of the same family (and thus, the electron, its neutrino, the $u$ and $d$ quarks and all their antiparticles should have the same constant assigned). This is a theoretical statement and it of course should also be subject to empirical tests, and indeed within the first family (largely the electron and the proton) it appears to be (unvoluntarily) tested in precision atomic physics, but it seems more interesting at the present time to devote some time to thinking whether $\hbar$ could take the same value for the second and third families as it does for the first.

To avoid much confusion, I will reserve $\hbar$ for the generic proportionality between momentum and the derivative of a field or wavefunction, but use $\hbar_a$ to denote the constant corresponding to each of the $a=$1, 2, 3 particle families, or specify $\hbar_\gamma$, $\hbar_\mu$, etc.

\section{Brief status of measurements and universality of Planck's constant}
\label{sec:measureh}

\subsection{$\hbar_\gamma\simeq \hbar_e$}

Let us start by Planck's constant for the photon, as obtained from Eq.~(\ref{photon}). While that equation was motivated by the study of black-body radiation, the most accurate early measurements of what I would call 
$h_\gamma$ are of course due to the photoelectric effect~\cite{Millikan}. 
Millikan obtained, at quoted 0.5\% error, 
$6.57\cd 10^{-34}$ Js, comparing well with Planck's $6.55$ estimate from blackbody radiation in the same units.
These were superseded by $X$-ray measurements~\cite{Steinerprog} where (circa 1962) 
\be \label{hphoton}
h_\gamma = 6.625\ 59(16)\times 10^{-34} {\rm Js}
\ee
a method that has been later abandoned to determine ``$h$'' since it involved measuring crystal spaces very precisely so they could be used to diffract X-rays and thus obtain their wavelength $\lambda$, to be compared with the electric $eV$ energy of the electrons accelerated to produce them by bremsstrahlung on a metal target. Still, they provide a crisp, accurate value of Planck's constant in its role in Eq.~(\ref{photon}) for the photon.

The standard, most accurate and promising method to measure $h$ at present time is to employ the relation
\be
h = \frac{4}{K_J^2 R_K}
\ee
that links Planck's constant with Josephson's constant $K_J=\frac{1}{\Phi_0}=\frac{2e}{h}$ and Von Klitzing's constant $R_K=\frac{h}{e^2}$.
The first yields the proportionality between the frequency of an induced AC current and its causing DC voltage at the insulating junction between two superconductors, $V = n \nu /K_J$; the second is the ratio of a quantum Hall resistance to the current intensity, $R_K=R_{\rm Hall}/I$~\cite{Stock}.

The dynamic Watt balance~\cite{Steiner,Kibble} allows to compare masses to electrical power. Its principle is the Lorentz force on the charge carriers when a wire is subject to a magnetic field, leading to a total force $F$ on the conducting wire of length $L$ carrying an intensity $I$ under a perpendicular field $B$ given by $F=BLI$. To avoid a direct measurement of the field $B$ and wire length $L$, an extra calibration step is taken.

Moving the wire with velocity $v$ through the magnetic field allows to eliminate the factor $BL$ since a potential $V_1=(BL)v$ builds up by Faraday induction, so that the balance equilibrium condition is
\be
mgv = V_1 I = V_1 V_2/R\ .
\ee
If the two electric potentials are calibrated against the Josephson effect and the resistance is taken from the quantum Hall effect,
\be
h=\frac{mgv}{{\rm const}f_1 f_2}
\ee
the measurement of $h$ has been reduced to that of careful weight, velocity, and circuit-frequency measurement.
After following the standard reasoning just presented, it is clear that the entire set of equations refers $h$ to the charge carriers in the wires, that is, $h_{\rm Watt} = h_e$.

The method yields, in SI units,
\be \label{helectron}
h_e = 6.626\ 068\ 89(23)\times 10^{-34}\ {\rm Js}
\ee
entailing a relative error $\frac{\Delta h_e}{h_e}\sim 3.4\cd 10^{-8}$.

In comparing Eq.~(\ref{hphoton}) and Eq.~(\ref{helectron}), I find a very satisfactory agreement 
\be \label{photonelectron}
\frac{\hbar_e-\hbar_\gamma}{\hbar_e} \simeq 72 (24) \times 10^{-6}\ ,
\ee
though short of the uncertainty in modern measurements of $\hbar_e$.

Other methods employed to measure $\hbar$ through accurate determinations of Avogadro's number such as an electrolysis measurement of Faraday's constant (the charge of $N_A$ electrons) boil down to an extraction of $\hbar_e$ and I will not discuss them further but refer to~\cite{codata}.

\subsection{Second and third fermion families}
A direct measurement of $\hbar$ for the particles of the second and third families would require simultaneously measuring both sides of Eq.~(\ref{deBroglie}) in some form. For example, one could think of measuring the width and lifetime of the same particle and use $\Gamma \tau=\hbar$.

But the best resolution that modern vertex detectors can reach is about
$c\tau\simeq 100\mu$m so that the characteristic energy scale is
$E\sim 2$ meV, whereas typical momentum resolution at a particle detector is 
$\delta p/p\sim 0.1$. If one thinks of typical decays of second and third family particles $K_s\to \pi\pi$, $J\psi\to n\pi$, $\tau\to \nu_\tau+n\pi$, etc. the pions or other particles in the final state have momenta in the 100 MeV range, so the precision in its determination is precise at the few MeV level: millielectronvolt precision is out of question.

Collider studies can achieve better accuracy when the beam energy can be carefully varied to scan a resonance, and for example the $J/\psi$ has a measured width of 93 keV~\cite{Beringer:1900zz}. But directly measuring its 2 picometer lifetime would require vertexing at the 10 picometer level (quite hopeless).

Thus, the capabilities to measure $\tau$ and $\Gamma$ are not commensurate, and both quantities cannot be obtained at the present time for the same particle. One needs to resort to indirect methods to constrain $\hbar$.

In the case of the second family, precision experiments on the muon (see the next subsec.~\ref{muon}) 
that provide one with two sources (if not more) of accurate data, the magnetic anomaly and the muonic Hydrogen atom,
may be used to constrain $\hbar_\mu$. 

Precision deteriorates for the third family. 
To impose some strong constraint on it I  resort to neutral instead of charged leptons.

A non-universal Planck's constant can make the kinetic term misaligned with the interaction term and thus the production (flavor) and propagation (kinetic) basis do not coincide, inducing mixing. I examine the case of neutrinos in section~\ref{sec:mixing}, though the phenomenon could equally well be discussed for neutral mesons or for any particles beyond the currently known Standard Model. 

Meanwhile, a test on lepton coupling universality is put to use in subsection~\ref{subsec:constantalpha}

\subsection{Muon properties}\label{muon}
 
The difference $\hbar_e-\hbar_\mu$  drops out of the muon's magnetic moment in leading order, before radiative corrections. 
 This comes about because the magnetic-moment anomaly is measured via the difference between the cyclotron (translational) and the spin precession frequencies~\cite{Bennett:2006fi}
\be
\omega_a = \omega_c-\omega_s = -a_\mu \frac{eB}{m_\mu}\ 
\ee
and $\hbar$ cancels between the left-hand side (quantization of muon orbits) and the right-hand side (muon magneton).

Turning to radiative corrections, because Planck's $\hbar$ enters the fine structure constant $\alpha:=\frac{e^2}{4\pi\hbar c}$, Schwinger's one-loop calculation of the magnetic anomaly
\be\label{schwinger}
\frac{g-2}{2} = \frac{\alpha}{2\pi}
\ee
is automatically sensitive to it.
This is the best measured property of the muon~\cite{Bennett:2006fi}, 
\be
\frac{g-2}{2}\ar_{\rm exp} = 0.001\ 165\ 920\ 89 (63)
\ee
and while a further experiment at Fermilab is in preparation, at present
the experimental uncertainty is
\be
\Delta \left( \frac{g-2}{2}\right) = 6\times 10^{-10} \ .
\ee 
Agreement with theoretical calculations is impressive and limited by the relatively low-precision of strong-interaction computations. 
A typical theory analysis would give
\be
\frac{g-2}{2}\ar_{\rm theory} = 0.001\ 165\ 918\ 27 (64)
\ee
and the discrepancy with the experimental measurement is approximately at the level of $3\times 10^9$, which is under intense discussion (as is the theory error~\cite{Goecke:2012qm,Williams:2013tia}).

Since the theory prediction at first order, Eq.~(\ref{schwinger}), is computed with the fine structure constant measured for the electron in atomic spectroscopy, the maximum discrepancy between theory and experiment for the muon's $g-2$ is an indirect test of universality of $\hbar$ in the scheme with universal charges $e_i\equiv e$. If I introduce a small quantity $\epsilon_\mu$ to parametrize deviations between $\hbar_e$ and $\hbar_\mu$ such that $\hbar_\mu = \hbar_e(1+\epsilon_\mu)$, 
then $\alpha_\mu = \alpha_e/(1+\epsilon_\mu)$ and Eq.~(\ref{schwinger})
\be
-\alpha_e \epsilon_\mu \simeq 2\pi \Delta \left( \frac{g-2}{2} \right)\ .
\ee
This leads directly to an upper limit for the possible variation of $\hbar$ between the first and second family,
\be \label{muonelectron}
\epsilon_\mu \simeq -2.6\cd 10^{-6}
\ee
which would, incidently, favor an $\hbar_\mu$  slightly smaller than $\hbar_e$ up to at most three parts per million.

A second problem in muon precision physics is the current outstanding~\cite{Pohl:2013yb} discrepancy between the calculated and measured Lamb shift in the muonic atom (conventional hydrogen with the electron substituted by a muon). The experiment at PSI results in  disagreement with calculations based on the standard proton radius, hence the name ``proton radius puzzle'', though in fact it might have nothing to do with the radius.
While fermion self-energy gives the main contribution to the Lamb shift in hydrogen, it is the photon vacuum polarization that is most important for the muonic atom, due to the tighter binding~\cite{Eides:2000xc,Pachucki:1996zza}, with
\ba
E(2P)-E(2S) & \simeq &0.0261789 \times \left(\frac{2}{3} \frac{m_{\mu{\ \rm reduced}} \ \alpha^3}{\pi}\right)\\ \nonumber
  &\simeq& 205.006\ {\rm meV} \ .
\ea
If again $\alpha_\mu =\alpha_e/(1+\epsilon_\mu)$, we need to replace 
$\alpha^3\to \alpha_\mu\alpha_e^2\to \alpha_e^3 (1-\epsilon_\mu)$ and,
\be
E(2P)-E(2S) \simeq 205.006\ {\rm meV} (1-\epsilon_\mu) \ .
\ee
The current disagreement between theory and experiment is $\Delta E_{\mu{\rm Lamb}}\simeq 320\mu$eV, with
\be
\frac{\Delta E_{\mu{\rm Lamb}}{\rm(exp-th)}}{E_{\mu{\rm Lamb}}} = 1.56\cd 10^{-3}\ .
\ee
The experimental figure is again larger than theory, so that putting the blame of the disagreement on $\hbar$ would lead again to a negative
$\epsilon_\mu\simeq -1.5\cd 10^{-3}$, leading to $\hbar_\mu<\hbar_e$.

As already noted by Jentschura in his analysis of millicharged particles~\cite{Jentschura:2010ha}, natural explanations of the Lamb-shift discrepancy are hardly compatible with the much tighter agreement between experimental and theoretical $g-2$.
In the case of $\hbar$, the magnetic anomaly leads to a maximum allowed variation of $\hbar$ between the muon and the electron that is a factor 250 smaller than required to fix the Lamb shift.

It is also worth remarking that there are further, not extremely accurate, flavor-physics tests; for example, Barger {\it et al.} have also found constraints on possible muon's anomalous couplings from kaon leptonic decays~\cite{Barger}.

To conclude this subsection, The Lamb shift of the muonic atom, and to a much larger extent the muon's magnetic anomaly, constrain the fine structure constant $\alpha_\mu$ to be close to $\alpha_e$, and thus also Planck's constant. The tighter constraint is due to $g-2$ and allows at most $\ar \hbar_\mu -\hbar_e\ar = O(10^{-6})$.

\subsection{Modifying $\hbar$ at constant coupling $\alpha$}\label{subsec:constantalpha}
Most of this work deals with modifications of $\hbar$ at fixed charge $e$, implying that
$\alpha_{\rm} = e^2/(4\pi\hbar c)$ also varies with $\hbar$ (and the same for other interactions). 
But now let me briefly address the case of a flavor-dependent $\hbar$ but with fixed coupling $\alpha_{\rm em}$ instead, that is, the non-universality of $\hbar$ is also present in the charges $e_i$ so that the coupling constant 
remains universal. 

Under this scenario many of the stringent tests weaken. For example, the $(g-2)$ anomaly of the muon's magnetic moment is much less constraining, since the perturbative series in $\alpha_{\rm em}$ of such success in the Standard Model remains valid.

A variation of $\hbar$ could nevertheless be seen in other places, for example in the Rydberg constant
\be \label{Rydbergdef}
R_\infty := \frac{m_e c^2 \alpha_{em}^2}{4\pi\hbar c} \ ,
\ee
so that, if again $\hbar_\mu = \hbar_e (1+\epsilon_\mu)$, the Rydberg constant extracted from the muonic atom would appear different from the Rydberg constant extracted from the electronic atom 
(the mass appearing is in both cases the mass of the electron) by
\be
R_\infty^{(\mu)} \simeq R_\infty^{(e)} (1-\epsilon_\mu) \ .
\ee

The PSI muonic atom experiment~\cite{Pohl:2010zza} discussed in subsection~\ref{muon}
would be consistent with QED precision computations based on 
conventional Hydrogen and proton structure studies if the Rydberg constant extracted from their experiment would be
$R_\infty^{(\mu)} = 10\ 973\ 731.568\ 160(16){\ \rm m}^{-1}$ as opposed to the CODATA~\cite{codata} value
$R_\infty^{(e)}   =  10\ 973\ 731.568\ 539(55){\ \rm m}^{-1}$, that represents almost a 5$\sigma$ shift. If interpreted as a difference of $\hbar$ at constant $\alpha_{\rm em}$, this would amount to 
$\epsilon_\mu = -3.45(65)\times 10^{-11}$.

Note that $\hbar_p\not = \hbar_e $ would also appear in the recoil corrections, since the proton mass in the non-relativistic hydrogen atom would effectively appear slightly different from that in the muonic atom because of the structure of the kinetic energy operator in Schr\"odinger's equation
\be\label{kineticS}
\left[
-\frac{\hbar_e^2}{2m_e}\left( \nabla^2_e +\left( \frac{\hbar_p^2}{\hbar_e^2}
\right) \frac{m_e}{m_p} \nabla^2_p \right) \psi(x_e,x_p) +V-E \right]
\psi(x_e,x_p) 
= 0
\ee
so that effectively, the nucleon mass measured in the two atoms would be slightly different $m_p\ar_{\mu H}\simeq m_p\ar_{H}(1+2\epsilon_\mu)$ (see the theoretical discussion on this point in subsection~\ref{subsec:composite} below).

The eigenvalues of the Dirac equation for a hydrogen-like atom (in the notation of Eq.(26) of~\cite{codata}), including recoil, are
\ba
E = Mc^2 + [f(n,j)-1] m_r c^2 - [f(n,j)-1]^2 \frac{m_r^2 c^2}{2M} +\dots \\ \nonumber
f(n,j) := \left(  1+\frac{(Z\alpha_{\rm em})^2}{(n-\delta)^2} \right)^{-1/2} \\ \nonumber
\delta := j+ 1/2 -\sqrt{(j+1/2)^2-(Z\alpha_{\rm em})^2}
\ea 
with $M$ the total and $m_r$ the reduced mass of the system. Since effectively, from Eq.~(\ref{kineticS})
$M_H=m_e+m_p(1-2\epsilon_p)$ and $M_{\mu H}=m_\mu + m_p (1+2(\epsilon_p-\epsilon_\mu))$,
the difference in binding energy for each level respect to the same without this non-standard recoil correction results in a slight shift of the muonic-atom levels not considered hitherto,
\be
\Delta E = \epsilon_\mu \left(
(f-1)m_r c^2 \frac{2m_\mu}{M}
-(f-1)^2 \frac{m_r^2c^2}{M^2}(2m_\mu-m_p)
\right)
\ee
that serves as an example on how a careful assessment of the entire Lamb shift computation 
would be necessary in a dedicated future study.

Now let us assess the third family briefly.
Again, it is best to focuse on its charged lepton, the $\tau$.
From the ratio (neutrino and antineutrino omitted) $\Gamma(\tau\to e)/\Gamma(\tau \to \mu)$, a test of universality follows~\cite{Pich2013} that yields, in modulus,
\be
\frac{g_\tau}{g_\mu} = 1.0006(21)
\ee
and from the ratio $\Gamma(\tau\to \mu)/\Gamma(\mu \to e)$, a second test is at hand
\be
\frac{g_\tau}{g_e} = 1.0024 (21) \ .
\ee
If instead of assessing charge universality I read these figures as tests of $\hbar$-universality for universal fine-structure constants $\alpha$, that is, interpret them as $\sqrt{\hbar_\tau/\hbar_{\mu}}=1.0006(21)$ and $\sqrt{\hbar_\tau/\hbar_{e}}=1.0024(21)$, the second of the two scenarios I considered, the following constraints follow,
\ba \label{fromtaudecay}
\hbar_\tau = \hbar_{\mu}\times 1.0012(42)\\
\hbar_\tau = \hbar_{e}\times 1.0048(42)
\ea
that have, in a quite obvious manner, a four per mille error, way insufficient to match the precision of the difference between the photon and the electron's, for example. 
No test of universality for the third family, e.g. the $\tau$ lepton, studied so far is as powerful as those for the muon, so I turn now to a distinct form of flavor mixing that can be induced by a non-universal $\hbar$.

\newpage
\section{Mixing by a flavor-dependent $\hbar$} \label{sec:mixing}
In this section I comment on $\hbar$-induced mixing through the kinetic term, an essential feature of the non-universality of $\hbar$.
\subsection{Basic formalism} \label{subsec:basicmixing}

The free, massless neutrino $\nu_i$ of species $i=1,2,3$,  would satisfy the Klein-Gordon equation (not summed over $i$)
$ \hbar_i^2 \Box \nu_i = 0$, 
with plane-wave solution (ignoring spin structure, irrelevant for the argument)  on the mass-shell $E=\ar {\bf p} \ar$
\be
\nu_i \propto e^{-i \frac{Et}{\hbar_i}} e^{i\frac{{\bf p}\cd{\bf x}}{\hbar_i}}
\ . 
\ee
At the origin ${\bf x}={\bf 0}$, the phase advance of a beam with well defined energy can simply be described by
\be \label{tderivative}
i\frac{d}{dt} \left( \begin{tabular}{c}
$\nu_1$ \\
$\nu_2$ \\
$\nu_3$ 
\end{tabular} \right) = \ar{\bf p}\ar
\left( \begin{tabular}{ccc}
$\frac{1}{\hbar_1}$& 0 & 0 \\
0 & $\frac{1}{\hbar_2}$& 0  \\
0 & 0 & $\frac{1}{\hbar_3}$\\
\end{tabular} \right)
\left( \begin{tabular}{c}
$\nu_1$ \\
$\nu_2$ \\
$\nu_3$ 
\end{tabular} \right) \ .
\ee
Now, the neutrinos would be emitted and observed as weak eigenstates that would not coincide with the propagation states of definite $\hbar_i$,
\be
\left( \begin{tabular}{c}
$\nu_e$ \\
$\nu_\mu$ \\
$\nu_\tau$ 
\end{tabular} \right) = {\mathcal{U}}
\left( \begin{tabular}{c}
$\nu_1$ \\
$\nu_2$ \\
$\nu_3$ 
\end{tabular} \right)
\ee
so that
\be
i\frac{d}{dt} \left( \begin{tabular}{c}
$\nu_e$ \\
$\nu_\mu$ \\
$\nu_\tau$ 
\end{tabular} \right)_L = \ar{\bf p}\ar {\mathcal{U}}
\left( \begin{tabular}{ccc}
$\frac{1}{\hbar_1}$& 0 & 0 \\
0 & $\frac{1}{\hbar_2}$& 0  \\
0 & 0 & $\frac{1}{\hbar_3}$\\
\end{tabular} \right)
{\mathcal{U}}^\da
\left( \begin{tabular}{c}
$\nu_e$ \\
$\nu_\mu$ \\
$\nu_\tau$ 
\end{tabular} \right)_L
\ee
or exponentiating,
\ba \label{mixing1}
\left( \begin{tabular}{c}
$\nu_e$ \\
$\nu_\mu$ \\
$\nu_\tau$ 
\end{tabular} \right)_L\!\! (t) &=&
\exp\left[
-i\ar{\bf p}\ar  {\mathcal{U}}
{\rm{diag}}\left(\hbar^{-1}_j\right)
 {\mathcal{U}}^\da
\right]
\left( \begin{tabular}{c}
$\nu_e$ \\
$\nu_\mu$ \\
$\nu_\tau$ 
\end{tabular} \right)_L\!\! (0) \\ \nonumber 
&=& {\mathcal{U}} \exp\left[
-i\ar{\bf p}\ar  
{\rm{diag}}\left(\hbar^{-1}_j\right)
 \right]{\mathcal{U}}^\da
\left( \begin{tabular}{c}
$\nu_e$ \\
$\nu_\mu$ \\
$\nu_\tau$ 
\end{tabular} \right)_L\!\! (0)
\ea
(the last step is easy to prove expanding both sides in a Taylor series and employing the unitarity of the flavor rotation matrix ${\mathcal{U}}{\mathcal{U}}^\da =I$).

Equation~(\ref{mixing1}) suffices to show flavor oscillations. If a neutrino was produced with definite flavor $\alpha$, the probability that it would have oscillated to flavor $\beta$ at a later time $t$ through the misalignment of the flavor basis with the $\hbar$-diagonal basis would be
\be
P\left(\nu_\alpha \to \nu_\beta\right) = 
\ar \sum_i U_{\beta i} e^{-i\ar{\bf p}\ar t/\hbar_i} U^*_{\alpha i}\ar^2
\ee
or pulling out a constant phase (irrelevant in view of the squared modulus), and defining 
$\Delta \hbar_i^{-1} \equiv \frac{1}{\hbar_i}-  \frac{1}{\hbar_1}$,
\be \label{mixing2}
P\left(\nu_\alpha \to \nu_\beta\right) = 
\ar \sum_i U_{\beta i} e^{-i\ar{\bf p}\ar t \Delta\hbar^{-1}_i} U^*_{\alpha i}\ar^2\ .
\ee
This is totally analogous to the usual neutrino-mass mixing formula~\cite{Morii:2004tp}. Perhaps it is worth showing the simplified case of two-family oscillations, where
\be
{\mathcal U} = \left(
\begin{tabular}{cc}
$\cos\theta$ & $\sin\theta$ \\
$-\sin\theta$ & $\cos\theta$
\end{tabular}
\right)
\ee
that reduces Eq.~(\ref{mixing2}) to the simple form
\be \label{oscillationshbar}
P\left(\nu_\alpha \to \nu_\beta\right)_\hbar = 
\left( \sin 2\theta \right)^2
\sin^2 \left(\frac{\ar {\bf p}\ar t}{2}\left(
\frac{1}{\hbar_2}-\frac{1}{\hbar_1}
\right) \right) \ .
\ee
Equation~(\ref{oscillationshbar}) with $E=\ar {\bf p}\ar$ 
is to be compared with the analogous equation for mass-induced neutrino oscillations with $E\simeq \ar {\bf p}\ar$, 
\be\label{oscillationmass}
P\left(\nu_\alpha \to \nu_\beta\right)_m =
\left( \sin 2\theta \right)^2
\sin^2 \left( \Delta m_{21}^2 \frac{t}{4E}
\right) \ .
\ee
The difference is clearly visible: while the advance of the flavor-oscillation phase in the conventional phenomenon induced by masses is
$\phi_m \propto \frac{1}{E}$, $\hbar$-induced oscillations see a phase advance
$\phi_\hbar \propto E$, so the two phenomena are easily distinguishable.

\subsection{Comparison with neutrino data}
\begin{figure}
\centerline{\includegraphics*[width=0.6\linewidth]{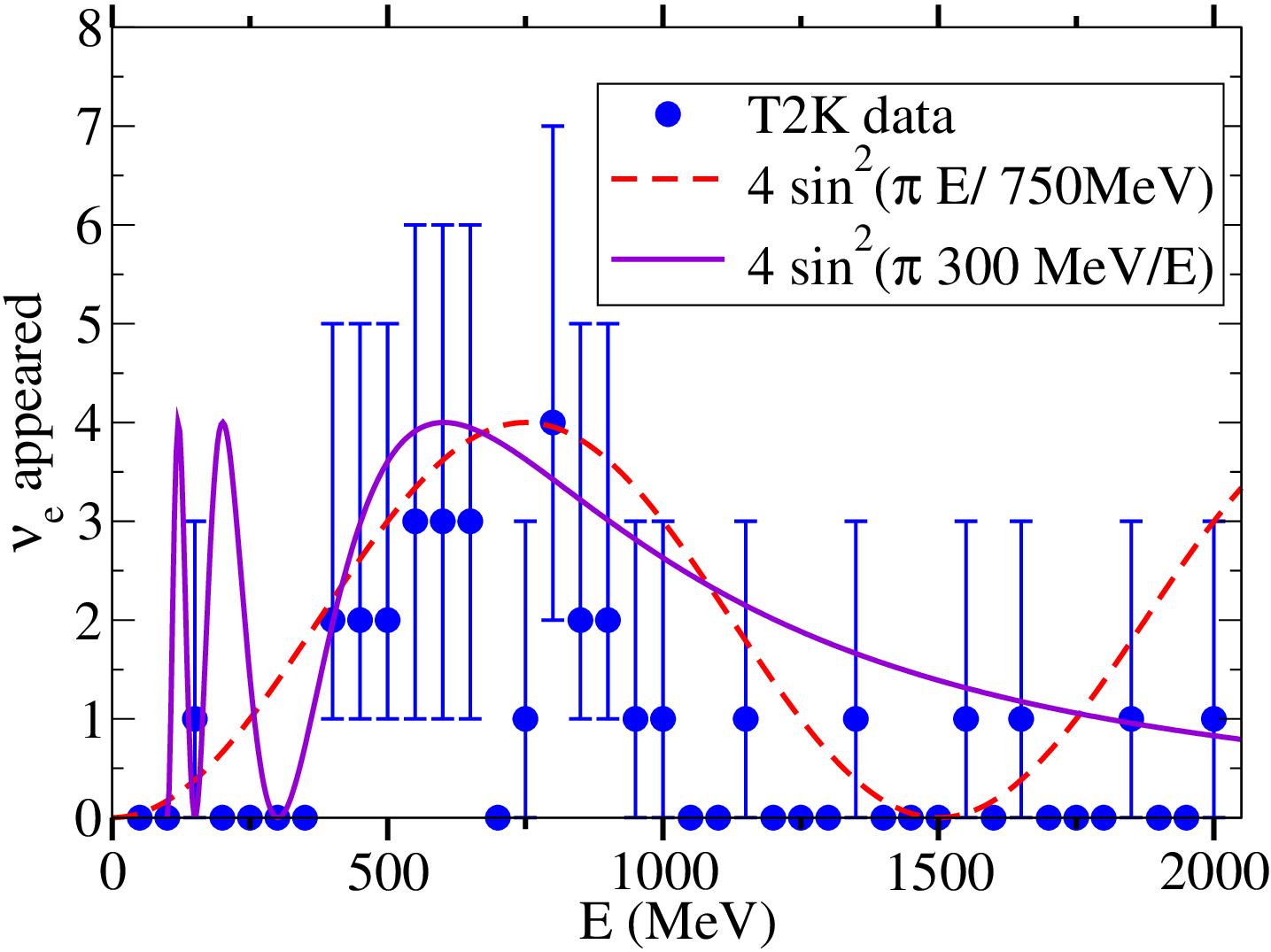}}
\caption{\label{fig:T2K}Appearance of electron neutrinos from an initial muon-neutrino beam as function of energy in the T2K experiment. The data is plot together with two ad-hoc functions with $1/E$ (standard; solid line) or $E$ ($\hbar$-induced; dashed line) phase advances.}
\end{figure}

I plot in figure~\ref{fig:T2K}  the recent data from the T2K collaboration~\cite{Abe:2013nka} in Japan, that prepares a $\nu_\mu$ beam and looks for the appearance of electron neutrinos $\nu_e$ downstream.
The data display an isolated maximum around 700 MeV, with very few events at the lowest and highest energies, so it cannot discriminate by itself which of the two mechanisms in Eq.~(\ref{oscillationshbar}) and Eq.~(\ref{oscillationmass}) is causing the flavor oscillations.

\begin{figure}
\centerline{\includegraphics*[width=0.6\linewidth]{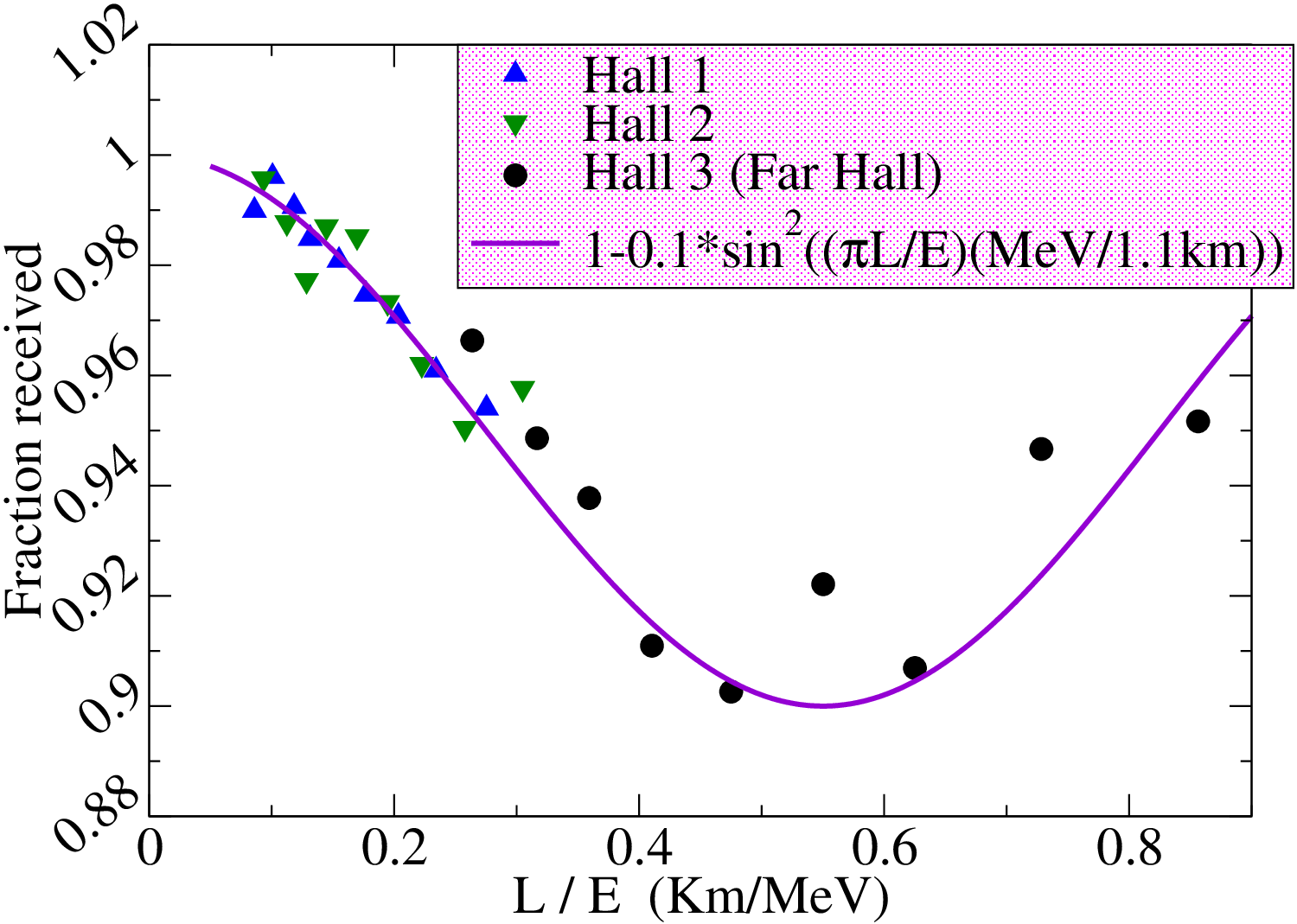}}
\centerline{\includegraphics*[width=0.6\linewidth]{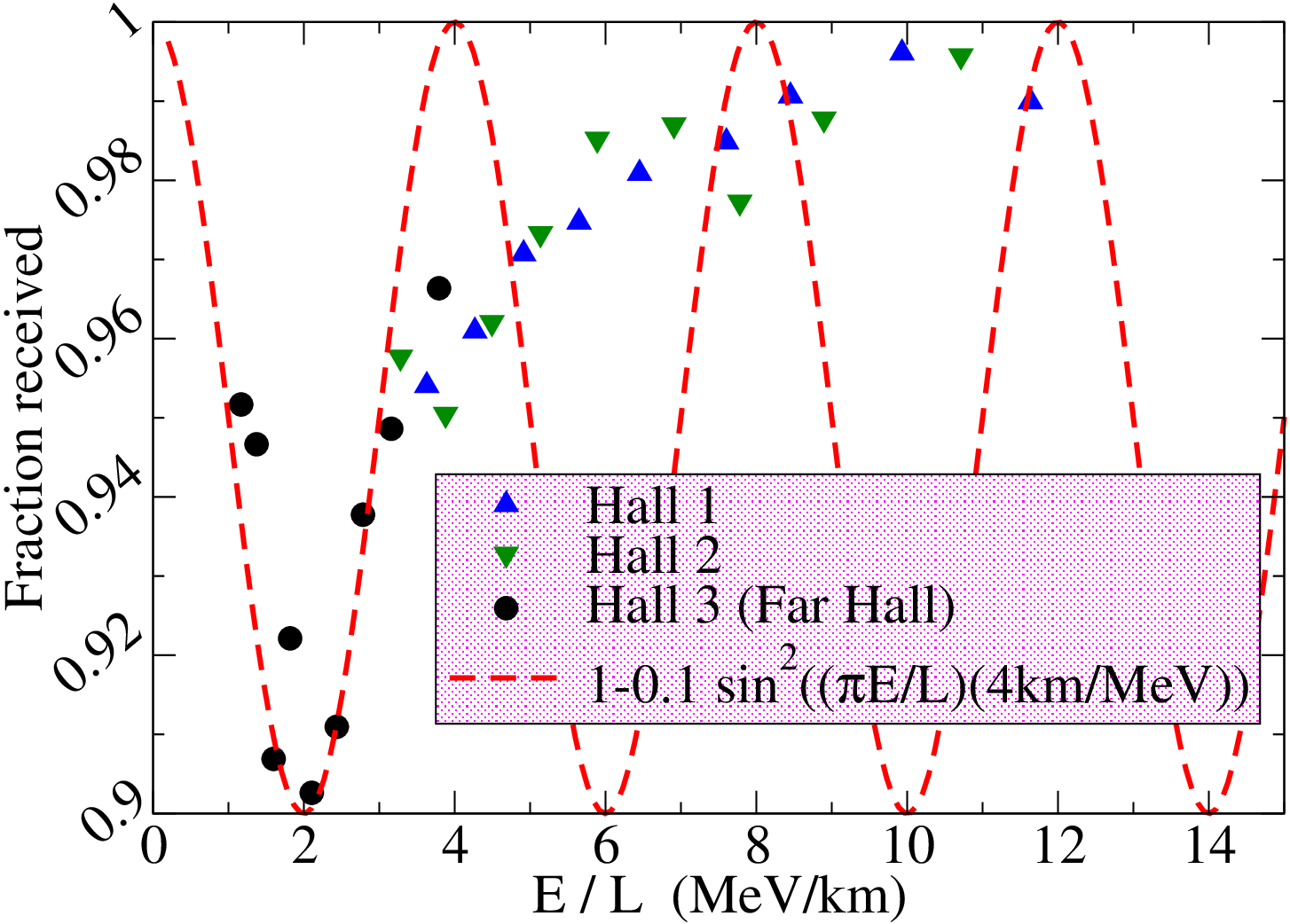}}
\caption{\label{fig:DayaBay} Data for reactor antineutrino disappearance 
at the three halls of the Daya Bay experiment. Top: plotted against $L/E$, as suggested by mass-driven oscillations. Bottom: plotted against $E/L$, as would result if $\hbar$ depended on the flavor and was the only effect present (constant factors are irrelevant). The data is totally conclusive and upholds the usual explanation with a phase advance of the flavor interference given by $\varphi\propto \Delta m_i^2 L / E$ instead of $\varphi\propto (Et/2)\Delta\hbar^{-1}$.}
\end{figure}

Turning to low-energy reactor neutrinos, I plot in figure~\ref{fig:DayaBay}
the recent spectrum provided by the Daya Bay collaboration~\cite{An:2013zwz}.
They provide separate data for the three experimental halls as function of $L/E$, the average distance of each hall to the active nuclear reactors, divided by the received antineutrino energy, and quote the fraction of antineutrinos received over those expected. This is directly replotted in the top graph of the figure, together with a squared sinusoidal shape that agrees roughly with the data. I do not display uncertainty bars in the figure for clarity and simplicity, since my purpose is not to provide a detailed fit of neutrino squared mass diferences; they can be tracked down in the Daya Bay publications. The data is nevertheless in reasonable agreement with the phase advance of the flavor interference in Eq.~(\ref{oscillationmass}), that should be proportional to $1/E$. 

To test the opposite hypothesis, whether the kinetic term (e.g. through $\hbar$'s non-universality) is causing the loss of emitted electron antineutrinos, it is best to plot the spectrum against $E$ instead of its inverse. To do it, one needs to know the average distance of the three experimental halls to the Ling-Ao and Daya Bay reactors, quoted in
~\cite{Dwyer:2013wqa} as $d_1=470$ m, $d_2=576$ m and $d_3=1648$ m (with the third hall being the furthest and perceiving the minimum of the oscillation; data taken there is represented by black circles in figure~\ref{fig:DayaBay}). 
The bottom plot of the figure clearly rejects the hypothesis that the phase advance might be linear in $E$ as there is one sharp minimum alone and the energy span is sufficient to have seen others, given its narrow width (hence, period of a putative periodic function also plotted to help the eye). 

One can safely conclude that a flavor-dependent $\hbar$ cannot drive the observed neutrino oscillations, as is naturally expected in view of the successful fits to neutrino data with existing theory. But now I can turn the question around to address the thrust of the investigation, asking what constraint on the variation of $\hbar$ is allowed by the neutrino data?
\subsection{Simultaneous $\Delta m$- and $\Delta \hbar$- induced mixing}
The question to answer in this subsection is whether, in the presence of neutrino masses as the conventional mixing mechanism, one can constrain the variation of $\hbar$. The equation equivalent to~(\ref{tderivative}) is now
(no sum over $i$)
\be
i\frac{d\nu_i}{dt} = \frac{\sqrt{p^2+m_i^2}}{\hbar_i} \nu_i
\ee
or, expanding the square root for $p\gg m_i$,
\be
i\frac{d\nu_i}{dt} \simeq \left(\frac{E}{\hbar_i}+
\frac{m_i^2}{2E\hbar} \right) \nu_i
\ee
where $\hbar$ is taken to be flavor-independent in the second term to avoid second-order infinitesimals. The two contributions just add up, and one easily obtains, in the simplified case of two-familly oscillation,
\be \label{2famosc}
P(\nu_\alpha\to \nu_{\beta\not = \alpha}) =
\sin^2 2\theta \sin^2\left(
\frac{Et(\Delta \hbar^{-1})}{2} + \frac{\Delta m^2}{2E\hbar}
\right) \ .
\ee

This form is ready for detailed fits of neutrino oscillations that have a small component of the phase advance proportional to the energy in addition to the larger piece with the inverse energy. There are no such fits for the newest neutrino data but older K2K combined with atmospheric neutrino oscillations, with a large energy span allowing to discern the energy functional form, have been studied for years~\cite{GonzalezGarcia:2004wg,GonzalezGarcia:2007ib} in search of deviations of the $1/E$ law that would point to BSM physics. New physics operators produce $E$, $E^2\dots$ terms in the lepton mixing phase, and I can invoke the studies constraining them to also bind possible interfamily deviations of $\hbar$.

Gonz\'alez-Garc\'{\i}a and Maltoni~\cite{GonzalezGarcia:2004wg,GonzalezGarcia:2007ib} considered (among other cases) the  oscillations driven by a Hamiltonian 
\ba \label{doubleosc}
H = \frac{\Delta m^2}{4E} \left( 
\begin{array}{cc} 
 -\cos 2\theta  & \sin 2\theta  \\  \sin 2\theta  & \cos 2\theta  \end{array} \right)
+\delta \frac{E}{2} \left( 
\begin{array}{cc} 
 -\cos 2\xi  & \sin 2\xi  \\  \sin 2\xi  & \cos 2\xi  \end{array} \right)
\ea
with $\theta$, $\xi$ two mixing angles left free for the fit, as would be $\Delta m^2$ and $\delta$.

The atmospheric neutrino energy extended up to energies of order 100 GeV and the data shows no sign of a phase increasing with energy. Therefrom, the authors conclude $\delta < 2\times 10^{-24})$ 
depending on several specific circumstances and the modulation by the possible mixing angle $\xi$.

This is a tiny number, so it is worth to see how such order of magnitude comes about. First, consider that the typical baseline scale in an atmospheric neutrino experiment is 2000km. Since no phase advance is seen, say to 0.1 rad, one would expect $ E L \delta <0.1 {\rm rad}$ and hence $\delta < 0.1\times 100 {\rm GeV} \times 2000{\rm km}/ (0.197{\rm GeV fm} )$ from where $\delta < 10^{-23}$.
Another way to arrive at it is by comparing the strength of the two terms in Eq.~(\ref{doubleosc}). Since the oscillations damp at high energy, the second term must be smaller than the first, and thus
\be
\delta < \frac{\Delta m^2}{2E^2}\ .
\ee
The typical neutrino squared mass diferences that fit this and other data are $\Delta m^2\sim 10^{-3} {\rm eV}^2$. For an energy of 100 GeV it follows that $\delta<10^{-25}$.
These two numbers bracket the result of the actual detailed fits, that is $\delta < 2\times 10^{-24})$ as mentioned. 

Translated to a variation of the Planck constant, by matching Eq.~(\ref{doubleosc})  and Eq.~(\ref{2famosc}),
\be
\frac{\Delta \hbar}{\hbar} = 2\delta
\ee
so that the bound obtained by Gonz\'alez-Garc\'{\i}a and Maltoni applies quite directly and I conclude that
\be \label{entreneutrinos}
\frac{\hbar_{\nu_\mu} - \hbar_{\nu_\tau}}{\hbar} < 10^{-24}
\ee
which is the tightest check of the universality of Planck's constant that I know of. No experiment in the laboratory can reach such precision in a direct or indirect measurement of $\hbar$ itself, but neutrino oscillations are a very sharp razor to examine the differences between constants for different particles. Incidently, as discussed below in subsection~\ref{sec:anomalies}, this is also a test of gauge invariance.

\section{$\hbar$ in the Lagrangian} \label{shufflehbar}
\subsection{Rescaling of the fields}
Planck's constant is usually omitted (because of the choice of natural units) from the QED Lagrangian density that describes a charged fermion flavor-multiplet field $\psi_a$ interacting with radiation $F_{\mu\nu}=A_{\nu,\mu}-A_{\mu,\nu}$ (the generalization to the weak interactions is straightforward and given shortly in Eq.~(\ref{LagWeak})),
\be
{\mathcal L}^{(1)}_{QED} = \sum_a \bar{\psi}_a \left( i\hbar_a \Dirac \partial + e\Dirac\! A -m_a\right) \psi_a - \frac{1}{4\hbar_\gamma} F_{\mu\nu} F^{\mu\nu} \ .
\ee
but I have avoided the usual choice of units $\hbar=1$ and kept Planck's constant explicitly while maintaining $c=1$ and a universal fermion charge $e$. Each of the terms has dimensions of energy per cubic length, $E/L^3$.

One can rescale the gauge field $A\to A' = \frac{A}{\sqrt{\hbar_\gamma}}$ to eliminate $\hbar_\gamma$ from the pure gauge term resulting in
\be
{\mathcal L}^{(2)}_{QED} = \sum_a \bar{\psi}_a \left( i\hbar_a \Dirac \partial
+e\sqrt{\hbar_\gamma} \Dirac\! A -m_a \right)\psi_a -\frac{1}{4} F_{\mu\nu}F^{\mu\nu} \ ,
\ee
with the dimensions of the vector field being now a square root of energy divided by length, $[A]=\sqrt{E/L}$ (instead of energy as in natural units).

A further rescaling of each fermion field $\psi \to \psi' = \sqrt{\hbar}_a \psi$ leads to
\be \label{Lag3}
{\mathcal L}^{(3)}_{QED} = \sum_a \bar{\psi}_a \left( i\Dirac \partial
+e\frac{\sqrt{\hbar_\gamma}}{\hbar_a} \Dirac\! A -\frac{m_a}{\hbar_a} \right)\psi_a -\frac{1}{4} F_{\mu\nu}F^{\mu\nu} \ .
\ee

In this form, the charge and mass of the fermion appear to absorb the entire effect of rescaling the fields; universality of the charge is broken if $\hbar$ depends on the particle, and any oscillation phenomenon (only apparently) seems driven by the mass term, since the kinetic term is now again proportional to the identity matrix in flavor space if there are several species, and thus diagonal in any basis.

But this last conclusion is too rushed. I of course wish to maintain the correct one-particle normalization in momentum space, 
\be \label{norm}
\la {\bf p}\lambda\ar {\bf q} \mu\ra = \hbar^3 (2\pi)^3\delta^{(3)}({\bf p}-{\bf q}) \delta_{\lambda \mu}\ .
\ee
In field terms, the one-particle state is obtained from the vacuum by a Fourier mode creation operator, whose normalization is fixed by the fact that there be exactly one particle,
\be
b^\da_\lambda({\bf p}) \ar 0\ra =\ar {\bf p},\lambda \ra\ .
\ee
The field rescaling needs therefore a $\sqrt{\hbar_a}$  to appear in the normal mode expansion; for example the charge-destruction part should read
\be
\psi^{(-)}_a = \sqrt{\hbar_a} \int\frac{d^3p}{(2\pi)^3\hbar_a^3\sqrt{2E_p}}
\sum_\lambda b_{\lambda}({\bf p}) u_\lambda({\bf p}) e^{ipx/\hbar_a} 
\ee
in textbook notation~\cite{Peskin:1995ev}.

But the anticommutation relations change due to the field rescaling, instead of 
\be
\{\psi_a({\bf x}),\psi_b({\bf y}) \} =\delta_{ab} \delta^{(3)}({\bf x}-{\bf y})
\ee
I now have one more power of $\hbar$ (if Planck's constant depends on flavor, then I need to specify $\hbar_a$),
\be \label{anticomm}
\{\psi'_a({\bf x}),\psi'_b({\bf y})\} =\hbar_a \delta_{ab} \delta^{(3)}({\bf x}-{\bf y})\ .
\ee

Leaving out the mass term, the weak Lagrangian reads, in analogy with Eq.~(\ref{Lag3}),
\be \label{LagWeak}
{\mathcal L}^{(3)}_{W} = \sum_a \bar{\psi}_a \left( i\delta_{ab}\Dirac \partial + g_{ab} \sqrt{\hbar_W}/\hbar_a \Dirac\! {\bf W} \right)\psi_b 
-\frac{1}{4} F_{\mu\nu}F^{\mu\nu} \ .
\ee

With this $\hbar$-rescaling of the gauge field the kinetic energy term in Eq.~(\ref{LagWeak}) remains diagonal under any unitary transformation, since it is proportional to the identity in flavor space, so one can use this liberty to diagonalize the weak interaction matrix $g_{ab}\to U g_{ab} U^\da \equiv diag(g_1,g_2,g_3)$. 
Apparently the flavor-oscillation phenomenon has disappeared since both terms are now diagonal. But it is Eq.~(\ref{anticomm}) that is now off-diagonal, 
\be \label{anticomm2}
\{\psi'_a({\bf x}),\psi'_b({\bf y})\} =
\left[ U diag(\hbar_1,\hbar_2,\hbar_3)U^\da \right]_{ab}
\delta^{(3)}({\bf x}-{\bf y})\ ,
\ee
evading an old argument of~\cite{weinberg} since the propagator will also be off-diagonal, due to such anticommutation relations.
\be
\frac{i\delta_{ab}}{p^2+i\epsilon} \to \frac{i\left[ U diag(\hbar_1,\hbar_2,\hbar_3)U^\da \right]_{ab}}{p^2+i\epsilon}
\ee
so that flavor mixing occurs nevertheless as in section~\ref{subsec:basicmixing}. 

In conclusion of this subsection, if $\hbar_a$ depends on flavor for fixed, universal charges $e_a=e$, then coupling universality is not exact. A rescaling of the fields does not make the oscillation phenomenon go away because diagonalizing the kinetic ${\mathcal L}_{\rm kin}$ makes the anticommutation relations non-diagonal instead.

\subsection{Anomaly cancellations and gauge invariance}\label{sec:anomalies}
I now recall why anomaly cancellations in the Standard Model~\cite{dobadonaranja,Foot:1992ui} provide a strong theoretical hint that Planck's constant must be the same across each of the particle families, that is, $\hbar_e = \hbar_u = \hbar_d = \hbar_{\nu_e}$ and similarly for their corresponding antiparticles.
In terms of the hypercharges $y_L$ and $y_R$, the conditions for SM gauge anomaly cancellation (and thus, renormalizability) are
\ba \label{anomaly1}
\sum_{u,d} (y_L-y_R) &=& 0\\ \label{anomaly2}
\left(N_c \sum_{u,d} + \sum_{e,\nu_e} \right) y_L &=& 0 \\ \label{anomaly3}
\left(N_c \sum_{u,d} + \sum_{e,\nu_e} \right)\left(y_L^3-y_R^3\right)&=& 0 \ .
\ea
These three conditions are satisfied by the standard hypercharge assignments
$y_{uL} = y_{dL} = 1/6$, $y_{uR}=2/3$, $y_{dR}=-1/3$, $y_{eR}=-1$ and
$y_{\nu L}=y_{eL}=-1/2$. Now, the hypercharge interaction term in the SM Lagrangian density is
\be \label{hyperchargeLag}
{\mathcal L}_Y = ig' \sum_j \bar \psi_j \gamma_\mu B^\mu\left(y_{jL}P_L + y_{jR}P_R \right)  \psi_j\ .
\ee
If each $j$-fermion's interaction term had to absorb a factor $\frac{\sqrt{\hbar_{B}}}{\hbar_j}$ as in 
Eq.~(\ref{Lag3}), Eq.~(\ref{anomaly1}), (\ref{anomaly2}) and (\ref{anomaly3}) would cease being satisfied, so the electroweak theory would become inconsistent.

This strongly suggests, from theory alone, that in the scenario with $\hbar$ non-universal but $e$, $g_s$, etc. universal, the variation of $\hbar$ can only occur from family to family, so that there are at most three fermion $\hbar$ possibilities, e.g. $\hbar_1:=\hbar_e=\hbar_{\nu_e}=\hbar_u=\hbar_d$. To this we have to add the (possibly) different constant for each of the gauge bosons (in low-energy particle physics, $\hbar_\gamma$, $\hbar_g$, $\hbar_W$), so there are at most six constants to constrain and see if they are consistent with one and the same number.

The second scenario with non-universal charges $g'_j$ does not seem constrained since Eq.~(\ref{hyperchargeLag}) has to be written  as
\be \label{hyperchargeLag2}
{\mathcal L}_Y = i \sum_j g'_j \bar \psi_j \gamma_\mu B^\mu\left(y_{jL}P_L + y_{jR}P_R \right)  \psi_j
\ee
with $\hbar_j$, $g'_j$ such that it is $\alpha$ that is universal. Thus, in this scheme each particle in the family may in principle have a different associated $\hbar$.

Next let us turn to the flavor mixing between families, where the triangle anomalies are of no concern since they cancel on a family by family basis. The discussion is best carried out for an Abelian local symmetry since the additional indices only complicate the notation without adding any new element.
An infinitesimal  gauge transformation is
\ba \label{gaugetrf}
\psi_i \to \psi_i' =(I_{ij}+ig_{ij}\alpha(x))\psi_j \\ \nonumber
A^\mu \to A^\mu + \delta A^\mu \ .
\ea
What would be the usual covariant derivative with the additional family index and a non-diagonal $\hbar$,
\be 
D^\mu_{ij} = ({\rm diag}\ \hbar)_{ij}\partial^\mu + i g_{ij} A^\mu
\ee
under the infinitesimal gauge transformation in Eq.~(\ref{gaugetrf}) becomes
\ba \label{trfD}
D^\mu_{ij}\psi_j \to D^\mu_{ij}\psi_j + i^2 \alpha A^\mu g_{ij} g_{jl} \psi_l +
i\alpha ({\rm diag}\ \hbar)_{ij} g_{jl} \partial^\mu \psi_l \\ \nonumber
+ig_{ij}\delta A^\mu \psi_l + i ({\rm diag}\ \hbar)_{ij} g_{jl} (\partial^\mu \alpha) \psi_l \ .
\ea
For this to be the usual law of a covariant derivative, behaving as Eq.~(\ref{gaugetrf}),
\be
D^\mu_{ij}\psi_j  \to  (I_{ij}+ig_{ij}\alpha(x)) D^\mu_{jk}\psi_k\ ,
\ee
in the last term of the first line of Eq.~(\ref{trfD}), $(\rm diag \hbar)$ and $g$ need to commute, implying that they can be diagonalized in the same basis. The absence of neutrino mixing induced by $\hbar$ is therefore supporting gauge invariance. In addition, the second line of Eq.~(\ref{trfD}) would need to cancel.
This means that gauge invariance is equivalent to the matrix identity $g=({\rm diag} \hbar) g$. 
As a consequence, a discrepancy in $\hbar$ between different families introduces an explicit gauge-violating term~\cite{weinberg}. This means that the tiny bound in Eq.(\ref{entreneutrinos}) is also a very good empirical test of gauge invariance.

\subsection{$\hbar$ for a composite system}\label{subsec:composite}

Consider now a composite system made of several particles (not necessarily a
well-determined number of them, a case in point being the proton), to each of
which I assign a slightly different Planck constant $\hbar_i$. 

At low resolution, the system appears itself as a particle, though we may know that it is
not fundamental after all, but that at higher energy some structure may be revealed.
What Planck constant should one assign to the whole object to characterize its momentum
$- i \hbar\nabla$ as per Eq.~(\ref{deBroglie})?

A general answer is not obvious and probably deserves an additional field-theory investigation by itself, 
so I will be contempt showing the simplified answer for a non-relativistic few body system, with one caveat as follows. 
If I would adopt a strictly non-relativistic quantum mechanical Hamiltonian, the only appearance of Planck's constant would be in the kinetic energy term, $-\sum_i \frac{\hbar_i^2}{2m_i}\nabla^2$. 
It is then obvious that the ratios between different Planck constants can be absorbed in the masses $m_i$, since the only appearance of both of masses and $\hbar$ is in these terms, in the fixed quotient $\hbar^2/m$. 

This is no more the case if I consider the full relativistic kinetic energy $\sqrt{p_i^2+m_i^2}$ that, when expanded, contains terms with different ratios $m_i$, $\hbar_i^4/m_i^3$, etc. so that not all factors of $\hbar_i$ can be absorbed at the same time by the mass.
So, if for example $m_1$ and $m_2$ are measured by balancing energy-momentum conservation in the decay products of other heavier particles using the relativistic mass-energy interconversion, they are fixed by an absolute, external reference energy and cannot be rescaled within the two-body system. The simplest way to keep track of this is to just retain $m_1+m_2$ in the non-relativistic Hamiltonian. 

Therefore, the Hamiltonian of interest is (adding a third or more particles as convenient)
\be
H = m_1+ m_2 -\frac{\hbar_1^2}{2m_1} \nabla_1^2 -\frac{\hbar_2^2}{2m_2} \nabla_2^2 +V({\bf r}_1-{\bf r}_2)\ .
\ee
To exploit translation invariance, implying that the potential depends only on the relative separation, 
it is useful to change variables to the relative and cm coordinates, the second being now modified to facilitate separability,
\ba
{\bf r} &=& {\bf r}_1-{\bf r}_2 \\
{\bf R} &=& \frac{\frac{m_1}{\hbar_1^2}{\bf r}_1 +\frac{m_2}{\hbar_2^2}{\bf r}_2}{\frac{m_1}{\hbar_1^2}+\frac{m_2}{\hbar_2^2}} \ .
\ea
In terms of these variables, the Hamiltonian becomes
\be
H = m_1+ m_2 -\frac{1}{2\left(\frac{m_1}{\hbar_1^2}+\frac{m_2}{\hbar_2^2}\right)} \nabla_R^2 -
\frac{1}{2} \left( \frac{\hbar_1^2}{m_1}+\frac{\hbar_2^2}{m_2}
\right)\nabla_r^2 +V({\bf r})\ ,
\ee
that can be written in the expected form
\be
H=m_1+ m_2 -\frac{\hbar_R^2}{2 M} \nabla_R^2 -\frac{\hbar_r^2}{2m_r} \nabla_r^2   +V({\bf r})
\ee
with the standard total and reduced masses $M:=m_1+m_2$, \ $m_r^{-1}:=m_1^{-1}+m_2^{-1}$, at the price of two different Planck constants for the total motion of the system and for the relative motion of its parts,
\ba \label{composite}
\hbar_{R}^2 &:=& \frac{\hbar_1^2 \hbar_2^2 M}{ \hbar_1^2 m_2+ \hbar_2^2 m_1}  
\\ \label{relative} 
\hbar_{r}^2 &:=& \hbar_1^2 \frac{m_2}{M} + \hbar_2^2 \frac{m_1}{M} \ .
\ea

It is clear from Eq.~(\ref{composite}) that $\hbar_R$, answering the question posed in this subsection, is intermediate between $\hbar_1$ and $\hbar_2$, and different from either. For equal masses it is a ``reduced'' $\hbar$ itself, up to a factor 2. Adding a third particle, it provides the quark-model answer to the proton's 
$\hbar_p$ given that of the constituent quarks $\hbar_u$, $\hbar_d$ as
($M=2m_u+m_d$)
\be
\hbar_p^2 = \frac{\hbar_u^2 \hbar_d^2 M}{\hbar_u^2 m_d + 2 \hbar_d^2 m_u}\ .
\ee

Saliently, the value of $\hbar_r$ to be assigned to the relative coordinate, Eq.~(\ref{relative}), differs for conventional hydrogen and muonic hydrogen in a now transparent way
\ba
\hbar_{H}^2     &:=& \hbar_e^2 \frac{m_p}{M} + \hbar_p^2 \frac{m_e}{M} \\
\hbar_{\mu H}^2 &:=& \hbar_\mu^2 \frac{m_p}{M} + \hbar_p^2 \frac{m_\mu}{M} \\
\ea
that makes clear why the Rydberg constant in Eq.~(\ref{Rydbergdef})
can be slightly different for the two systems, thus providing an exotic explanation for the muonic atom discrepancy, should other stringent tests allow.

\section{Discussion and summary}
In this article I have considered whether Planck's constant, $\hbar$, takes the same value for the various fundamental particles in the standard model, and given some examples of constraints that come to mind. It is surprising that the issue has not been revisited for very long, particularly since after the confusion caused by the CERN to Gran Sasso superluminal-neutrino claim~\cite{opera} we have learnt that indeed the universality of the speed of light $c$ seems to be in good shape.

It is clear that if Planck's constant $\hbar$ was not universal, the fundamental unit of length constructed from $c$, $G_N$ and $\hbar$, Planck's scale $L_P:= \sqrt{\hbar G /c^3}$
at which quantum gravity effects become strong,
would also depend on the particle family. Happily, since the allowed variations of $\hbar$ seem to be small, the various Planck scales are near each other, and from our ``low-energy'' prospective there is not much of a difference. 

Planck's constant is determined at present with what one would call excellent accuracy; the Codata group~\cite{codata} quotes, considering several alternative measurements,
\be
h = 6.626\ 069\ 57(27)\times 10^{-34}\ {\rm Js}
\ee
which has a relative error of $4.4\times 10^{-8}$ and is dominated by $h_e$ measurements~\footnote{
There is in addition interest in redefining the kilogram~\cite{StockMetro}, basic mass quantity in the SI. The reason is that the weight differences between various Platinum-Iridium cylinders in use as standards worldwide are apparently drifting by as much as half a microgram per year. A definition of the kg based on accurate measurements of $\hbar$ require this quantity to be known to a precision of $10^{-8}$. Temptatively, I would expect such accurate upcoming measurements to continue being performed with the Watt balance and to correspond to $\hbar_e$ with $m_e$ providing the mass scale.}.
I have shown constraints to $\hbar_\gamma-\hbar_e$ in Eq.~(\ref{photonelectron}), $\hbar_\mu-\hbar_e$ in Eq.~(\ref{photonelectron}), $\hbar_\tau$ respect to the later two in Eq.~(\ref{fromtaudecay}), and as the sharpest test,
$\hbar_{\nu_\mu}-\hbar_{\nu_\tau}$ in Eq.~(\ref{entreneutrinos}), based on the observation that a non-universal $\hbar$ could cause non-standard oscillations.

Additional sharp tests on leptons could follow if, for example, the muonium atom $\mu^+\mu^-$ could be produced~\cite{Brodsky:2009gx}.

Is there sufficient cause to suspect that $\hbar_i\not =\hbar_j$ for some two particles? Not really at this time. Is there sufficient cause to investigate it further? Yes indeed, since some experiments, the Lamb shift in the muonic atom, for example, are puzzles awaiting a solution that defies theory. 

I have barely touched on the issue of the strong interactions, that would hide
$\hbar_g$ and $\hbar_q$ behind hadron structure, though I dedicated subsection
~\ref{subsec:composite} to elementary considerations on compositeness. How to access the $\hbar$ factors of colored particles with accuracy is worth thinking about, probably in terms of high-energy processes and perturbation theory.

I should also perhaps trivially remark that the often read statement ``the observation of neutrino oscillations implies that neutrinos have mass'' implicitly implies that gauge invariance is exact and that $\hbar$ is universal, otherwise the kinetic term can cause flavor mixing. What makes the standard scenario undoubtable is the spectacular agreement reached in fitting many neutrino energy spectra.



I have only passed with a thick brush over several important topics that surface when examining a non-universal $\hbar$, but the interested reader can delve in many phenomena that it would be interesting to revisit. 

\acknowledgments
I thank C. Gonz\'alez-Garc\'{\i}a and Oscar Vives for patient comments.\\
Work supported by spanish grant FPA2011-27853-C02-01 and CPAN Consolider-Ingenio.\\



\end{document}